\documentclass[sigconf,authorversion]{acmart}

\usepackage{hyperref}
\usepackage{libertine}
\usepackage{graphicx}
\usepackage{subfig}  
\usepackage{balance}  
\usepackage{enumitem}
\usepackage{listings}
\usepackage{amsmath}
\usepackage{stfloats}
\usepackage{latexsym} 
\usepackage{url}
\usepackage[textsize=tiny]{todonotes}
\usepackage{xcolor}
\usepackage{tabularx}
\usepackage{booktabs}
\usepackage{color,soul}
\usepackage{framed}
\usepackage{verbatimbox,lipsum}
\usepackage{tikz}
\usepackage{float}
\usetikzlibrary{arrows}
\usetikzlibrary{shapes}

\usepackage[ruled,vlined,linesnumbered,noresetcount]{algorithm2e}
\usepackage{varwidth}

\usepackage{mathtools}

\copyrightyear{2025} 
\acmYear{2025} 
\setcopyright{acmlicensed}\acmConference[to appear in ADBIS]{ADBIS2025]to appear in ADBIS }{2025}{}
\acmBooktitle{.}
\acmPrice{.}
\acmDOI{}
\acmISBN{}

\begin{document}

% *** TITLE ****************************************
\title{PUL: Pre-load in Software for Caches Wouldn't Always Play Along}

% *** AUTHORS **************************************
\author{
Arthur Bernhardt$^{*}$, 
Sajjad Tamimi$^{\#}$, 
Florian Stock$^{\#}$, 
Andreas Koch$^{\#}$, 
Ilia Petrov$^{*}$
}
\affiliation{%
	\institution{
	$^{\#}$Embedded Systems and Applications Group, 
	$^{*}$Data Management Lab
	}
	\institution{
	$^{\#}$Technische Universit\"at Darmstadt, 
	$^{*}$Reutlingen University
	}
	\country{} % BERNHARA: Fixes Compile Errors
}

% *** COMMANDS **************************************
\renewcommand{\sf}{\sffamily}
\newcommand{\nKVtitle}{\textsf{\textbf{\lowercase{\underline{n}}KV}}}
\newcommand{\nKVcaption}{\textsf{\textbf{nKV}}}
\newcommand{\nKV}{{\sf \lowercase{\underline{n}}KV}}
\newcommand{\PUL}{\textsf{PUL}}
\newcommand{\neoDB}{{\sf \lowercase{neo}DBMS}}
\newcommand{\pimDB}{{\sf \lowercase{pim}DB}}
\newcommand{\upmem}{UPMEM}
\renewcommand{\shorttitle}{}
\renewcommand{\shortauthors}{A. Bernhardt et al.}
\newcommand\circlearound[1]{% 
  \tikz[baseline]\node[draw,shape=circle,anchor=base,minimum size=10pt,inner sep=0pt] {#1} ;}

%*****************************************
\begin{abstract}
Memory latencies and bandwidth are major factors, limiting system performance and scalability. Modern CPUs aim  at \textit{hiding} latencies by employing large caches, out-of-order execution, or complex hardware prefetchers. However, software-based prefetching exhibits higher efficiency, improving  with newer CPU generations. 

In this paper we investigate software-based, post-Moore systems that offload operations to intelligent memories. We show that software-based prefetching has even higher potential in near-data processing settings by maximizing compute utilization through compute/IO interleaving. 
\end{abstract}

    \maketitle
%******************************************

\section{Introduction}
\label{sect:intro}
%******************************************
\noindent\textbf{Memory latencies and bandwidth are major factors, limiting system performance and scalability.} 
%------------------------------------------
Memory is getting colder as over the last 20 years DRAM capacity improved 128$\times$, bandwidth 20$\times$, latencies only 1.3$\times$ \cite{Mutlu:MECO:2018} and it takes several hundred nanoseconds to access it \cite{MemLatency:2022,Teubner:DataProcFPGA:2013}.
To address the so called \textit{memory wall} modern CPU hardware aims at \textit{hiding} latencies by employing large caches, out-of-order execution, and complex hardware prefetchers. Unfortunately, they quickly reach their limits on non-predictable access patterns. Aggravating matters, data-intensive systems face low cache-efficiency due to low data locality, and extensive data transfers of growing datasets.

\noindent\textbf{Software-based prefetching presents a promising direction delivering steadily improving results with newer hardware generations.} 
%------------------------------------------
Software prefetching is a well-studied technique \cite{Shimin:HashJoinPrefetching:ICDE:2004,Shimin:IndexPrefetching:SIGMOD:2001,Falsafi:AccessChaining:VLDB:2015,Minhas:Coroutines:VLDB:2018,Kohler:CacheCarftiness:EuroSys:2012,Teubner:MxTasks:SIGMOD:2021,Psaropoulos:Interleaving:VLDB:2017,Vuduc:Prefetching:TACO:2012} that allows software to provide hints to the CPU data prefetcher about future data accesses and hard to predict patterns using special instructions. Recent study by Kühn et al. \cite{Teubner:Prefetching:DAMON:2023} thoroughly examines the merits and caveats of software prefetching on CPUs, indicating that its efficiency improves with each of the newer hardware generations. Notably, by hinting at upcoming accesses software prefetching becomes most effective if coupled to data structures, their layouts, operations and cache-optimisations \cite{Shimin:HashJoinPrefetching:ICDE:2004, Shimin:IndexPrefetching:SIGMOD:2001, Falsafi:AccessChaining:VLDB:2015, Minhas:Coroutines:VLDB:2018, Teubner:MxTasks:SIGMOD:2021, Psaropoulos:Interleaving:VLDB:2017}. However, CPU-based software prefetching does not cover the complete problem space.

\noindent\textbf{Novel Post-Moore architectures introduce additional challenges.} 
%------------------------------------------
Modern post-Moore architectures take a different avenue to mitigate the impact of high latencies and bandwidth limitations, by attempting to offload data processing operations to some kind of processing element (PE) located close to or even within the physical memory. As a result Near-Data Processing (NDP) becomes possible \textit{in-memory} (as so called Processing-in-Memory -- PIM), \textit{in-storage} (on smart/computational storage), or i\textit{n-network} (on SmartNICs). To remain \textit{economical} and comply with the \textit{commodity model, under which memory and storage are sold} these are simple in-order PEs that are \textit{slow} (150 -- 500 MHz) and with \textit{cache and memory hierarchies} different than those on CPUs. Such intelligent memories/devices typically operate on various memories with a broad latency spectrum (e.g., FLASH, NVM, DRAM, HBM, SRAM). 
What's more, the in-situ \textit{cache/memory organisation differs} from those on CPUs and involves scratchpad memories instead of caches, therefore well-known cache-optimisations do not work, even if host-CPUs and in-situ PEs operate on the same data and layouts.  
Furthermore, the weak PEs get slowed down further by high memory latencies which, depending on the underlying memory technologies, may have a wide range. On the upside, there is ample bandwidth and throughput in-situ. 

Therefore software prefetching emerges as a technique for \textit{maximizing the compute utilisation on the weak PEs by interleaving compute and I/O}.

\begin{figure}[t]
	\begin{center}
         \includegraphics[width=\columnwidth]{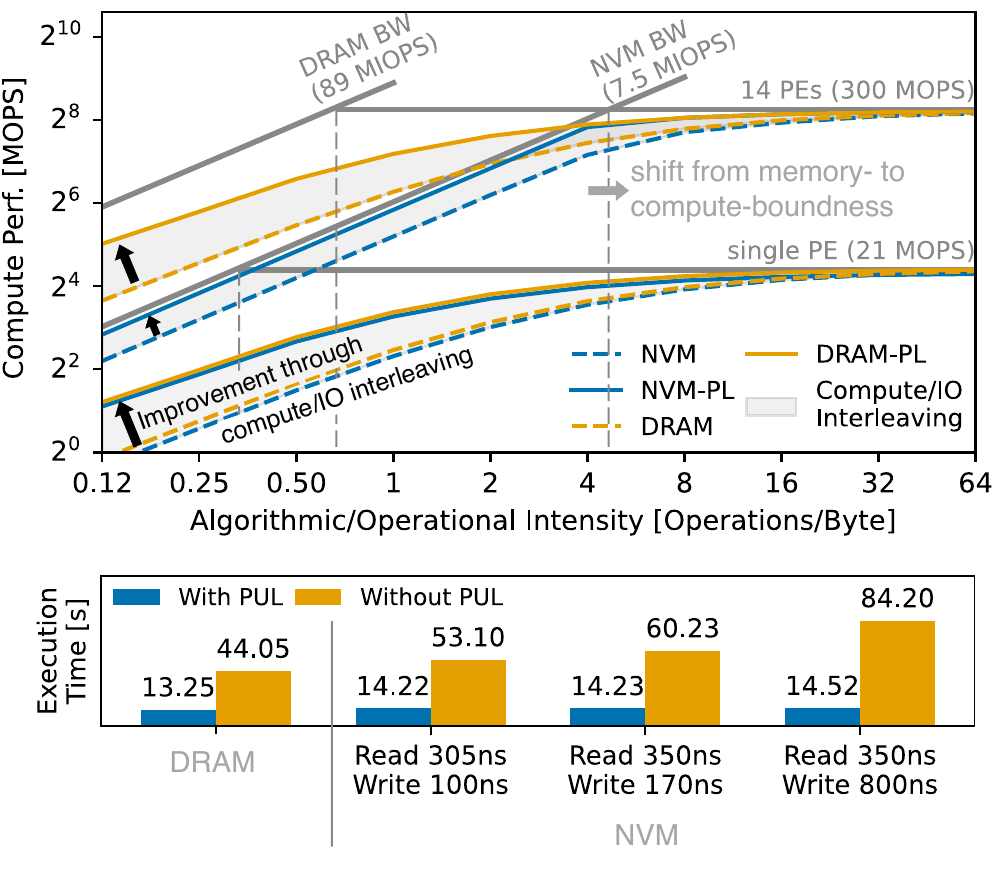}
         \vspace{-10pt}
         \caption{(top) Roofline analysis showing the potential for pre-loading. \PUL{} increases compute utilisation $\geq2\times$ through compute/IO interleaving with low algorithmic intensities on DRAM and NVM in NDP settings. (bottom) Mitigation of the impact of higher latencies through \PUL{}.}
	\label{fig:headpicture}
	\end{center} 
    \vspace{-10pt}
\end{figure} 

\noindent\textbf{\PUL{}: custom software \underline{P}re- and \underline{U}n-\underline{L}oading to the rescue.}
%------------------------------------------
In this paper, we introduce novel preloading techniques that facilitate interleaving compute and data transfers and leverage PE's low compute power by better utilizing the in-situ scratchpads. Preloading enables byte-addressable asynchronous application-defined transfers from arbitrary addresses into small PE-attached scratchpad memories (SRAM), while the PE performs useful work. In the mean time, persisting temporary or final PE results suffers from the inverse problem, as the PE is slowed down by expensive flushes from scratchpads to persistent storage. To this end, we employ \textit{unloading} in combination with scratchpad double-buffering strategies to asynchronously write out results. 

We distinguish between software prefetching/flushing and preloading/unloading. The former is inherently tied to CPU-specific cache hierarchies as well as vendor-defined instructions to hint at fixed-size cache-line transfers in advance without explicit control over data cache-placement or cache-eviction. The latter (focus of the present paper) employs externally attached SRAM-based "cache-like" scratchpad memory with explicitly managed, variable-sized data transfers. Both mechanisms require clear differentiation as they are not mutually exclusive.

\noindent \textbf{Our initial experiment}
%------------------------------------------
demonstrates the effectiveness of \PUL{} for utilizing the compute/IO interleaving (Fig. \ref{fig:headpicture}). To this end we investigate the different algorithmic/operational intensities for DRAM and slow NVM with different numbers of PEs. The compute/IO interleaving is especially prominent for typical DBMS operations that tend to have low operational intensity \cite{Mutlu:BenmarkingNewParadigm:IEEEAccess:2022}. Furthermore, \PUL{}-based software preloading and unloading enables hiding high NVM latencies and improving bandwidth utilisation (Fig. \ref{fig:headpicture}, bottom).

\noindent The \textbf{contributions} of this paper are:
%------------------------------------------

\begin{itemize}[leftmargin=*,noitemsep,nolistsep,nosep]
	\item 
	We investigate the effectiveness of \PUL{} in novel, post-Moore architectures 
    such as PIM and NDP, exploring the potential for compute and I/O interleaving on real hardware.

    \item
    We evaluate the impact of different memory latencies (e.g., DRAM, NVM) on performance and efficiency of compute/IO interleaving strategies and provide insights for further optimizations.

    \item
    We introduce \textit{unloading} to complement preloading and prefetching techniques, providing a new opportunity for optimizing compute performance and enhanced system efficiency. 
 
	\item 
    We demonstrate a novel DMA-engine, specifically tailored for \PUL{} and present insights in designing, optimizing, and employing such custom DMA-engine in NDP.    

\end{itemize}

\section{PUL Architectures and Interfaces}
\label{sect:pul}
%***************************************** 
One of the most noticeable differences between traditional CPU-architectures and emerging paradigms like PIM and NDP lies in their raw compute capabilities per core. In contrast to recent CPU systems, which already reach clock speeds of 4-6 GHz, commercially available PIM systems like the PIM-enabled memory DIMMs from UPMEM only reach 350 MHz and are expected to increase to 466 MHz in later generations. Furthermore, recently published architectures for NDP on Field-Programmable Gate Arrays (FPGAs) \cite{ Bernhard:neoDB:ICDE:2022, Vincon:UpdateAwareNDP:vldb:2022,Tamimi:TRETS2024:DANSEN} have demonstrated PE clock frequency rates as low as 150MHz. This disparity highlights the significance of maximizing the utilization of all available clock cycles in such systems.

\begin{figure}[b] 
	\begin{center}
         \includegraphics[width=\columnwidth]{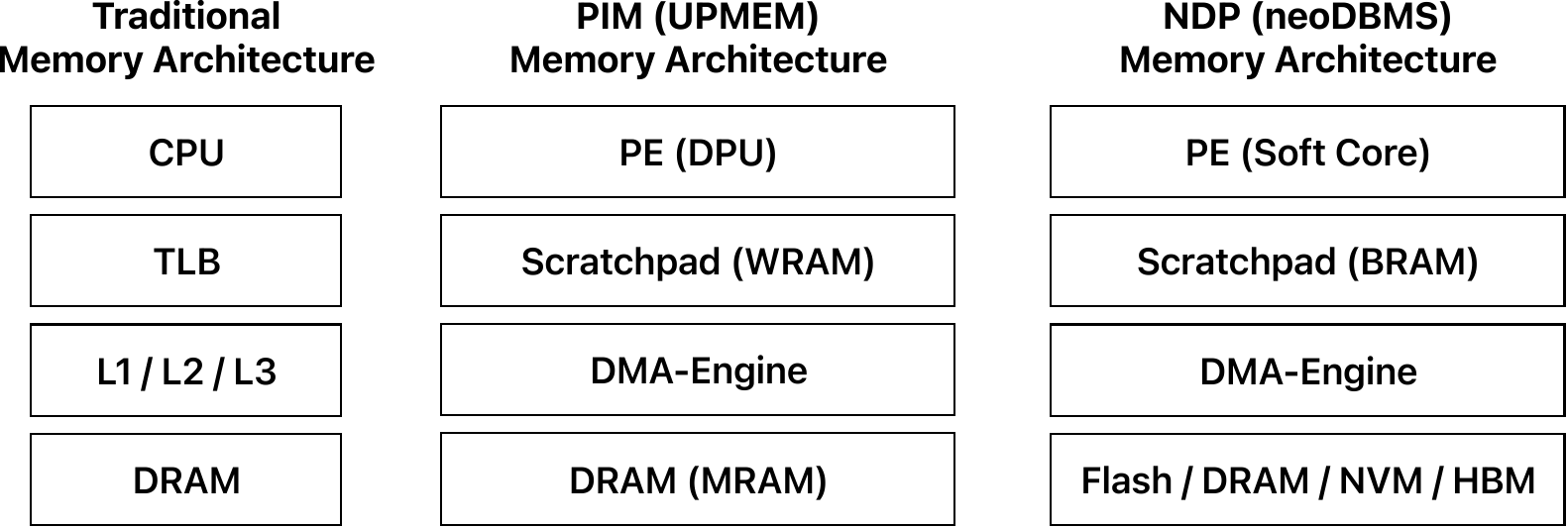}
		\caption{Characteristics of Post-Moore architectures differ greatly from traditional CPU-based architectures.}
	\label{fig:architectures}
	\end{center} 
\end{figure}

PEs are designed to be resource-conscious, featuring in-order operation and lean memory hierarchies without multiple cache levels or complex HW prefetching mechanism found in traditional CPU architectures (Fig. \ref{fig:architectures}). This simplicity enables the dense packaging of a large number of PE units within a single device. For instance, UPMEM supports up to 128 PEs (called DPU) per DIMM\cite{upmem:hotchips:2019}. The high degree of parallelism is facilitated by the systems being inherently non-cache-coherent. This does present its own challenges in terms of programming model and synchronization, but is considered out of scope for this paper. However, the benefits of reduced cache contention, particularly in async. preloading/unloading and managing cached data, make this design choice highly economical.

\begin{lstlisting}[
           language=C,
           showspaces=false,
           basicstyle=\ttfamily\smaller,
           keywordstyle=\bfseries,
           numbers=left,
           numberstyle=\small,
           commentstyle=\color{blue},
            stringstyle=\color{gray},
            xleftmargin=0.5cm,
            morekeywords={PRELOAD,UNLOAD,PRELOAD_WAIT,PRELOAD_SET_SIZE},
           caption={Example of software pre- and un-loading in \PUL{}.},
           label=lst:preloading
        ]
PRELOAD_SET_SIZE(64);
...
PRELOAD_WAIT();                     // Synchronize 
// Begin I/O and Compute Interleaving ============
// Async. I/O
PRELOAD(rand_ptr[4], bram_ptr[4]); // ---
PRELOAD(rand_ptr[5], bram_ptr[5]); //   | Preload-
PRELOAD(rand_ptr[6], bram_ptr[6]); //   | Distance
PRELOAD(rand_ptr[7], bram_ptr[7]); //   | of 4
// Interleaved Compute                  |
result += bram_ptr[0];  // <-------------    
result += bram_ptr[1];      
result += bram_ptr[2];      
result += bram_ptr[3];      
// Async. I/O
UNLOAD(bram_ptr[0], nvm_ptr, 64*4);  
// End Interleaving ==============================
PRELOAD_WAIT();                     // Synchronize 
\end{lstlisting}

\PUL{} leverages architectures that employ explicit memory management as in PIM or NDP, where software takes full control over memory access and caching. Data requests are directly dispatched to DMA engines, which asynchronously transfer data to fast, cache-like scratchpad memories, allowing PEs to access it in a single cycle. Notably, these systems do not rely on the traditional virtual memory abstraction, instead performing direct access to physical memory addresses. This allows for more efficient request processing, as the system can bypass intermediate layers of abstraction and operate at the hardware level. For example, TLB misses on CPU-based architectures and SW-prefetching pose an considerable overhead \cite{Teubner:Prefetching:DAMON:2023}, requiring applications to utilize \textit{huge pages}, physically contiguous memory allocated in advance, to mitigate the issue.

To fully leverage the benefits of \PUL{}, it is crucial to interleave compute and I/O operations seamlessly. This mandates that the DMA engine concurrently accepts new requests, while simultaneously processing already dispatched ones, without blocking the PE. Furthermore, the PE must be able to continue processing operations without waiting for the DMA-transfer from the local DRAM to the PE-local scratchpad to complete. Prior works \cite{Bernhardt:pimDB:damon, prim} utilize UPMEM's PEs to leverage \PUL{} through a programming model that employs software kernels and multiple Tasklets. Tasklets are lightweight software threads in the UPMEM programming model that can be executed in parallel on a single PE (UPMEM DPU). By dispatching DMA requests on some Tasklets, while other Tasklets concurrently utilize clock cycles, the PE interleaves I/O and compute operations.  

Devices built upon FPGAs allow a high degree of freedom on HW design, including custom memory hierarchies. We leverage this flexibility in our smart storage devices (Fig. \ref{fig:architectures}-NDP), which employ soft-core PEs for in-situ processing, by designing and deploying a custom DMA engine specifically for \PUL{}. This custom DMA-engine, bundled with fast scratchpad memory, e.g. FPGA BRAM, allows PEs to efficiently manage data transfer and compute. Through a simple programming interface (Listing \ref{lst:preloading}), the PE can dispatch both DMA preload and unload requests to the custom DMA-engine by writing physical source and destination addresses, as well as transfer sizes, into exposed HW registers. The custom DMA-engine features two FIFO queues, one for preload and one for unload requests, each capable of storing up to 64 requests simultaneously. Synchronizing the asynchronous non-blocking DMA requests is achieved by reading from an exposed HW status-register. This register returns the status of the FIFO, indicating whether all requests have completed or if some remain pending in the queue. While this simple synchronization approach has proven sufficient for \PUL{}, as preload and unload requests can already be synchronized independently, we are still investigating the potential benefits of a more fine-grained FIFO status handling, i.e. identifying specific pending requests, to further optimize \PUL{} efficiency. Additional optimization techniques employed in our design include register value buffering. When DMA requests do not alter register values, for example, the transfer size remains the same, we can skip writing to the register altogether. As an external module, our \PUL{} HW does not assume any specific characteristics of the underlying PEs or HW.

\noindent\textbf{Factors impacting the quality of preloading.}
%------------------------------------------
 Most importantly, data structures/layouts and algorithms are designed and optimized for CPU-architectures and caches, where prefetching operates at fix-sized cacheline transfers. Configurable transfer sizes in \PUL{} present opportunities to explore new strategies in layout and access-based optimizations. Similarly, data structures have direct impact on the execution models and might pose constraints on gathering and firing sufficient requests in advance. Proposed solutions like grouping \cite{Shimin:HashJoinPrefetching:ICDE:2004, Falsafi:AccessChaining:VLDB:2015}, coroutines \cite{corobase, corograph, Minhas:Coroutines:VLDB:2018, Psaropoulos:Interleaving:VLDB:2017}, or fine-grained tasks \cite{Teubner:SIGMOD:MxTasks:2021}, have yet to be reevaluated for such new memory architectures that do not run the risk of premature cache-evictions, or require deliberate prefetch-timing as discussed in \cite{aptget}.

\section{Experiments and insights}
\label{sect:eval}
%***************************************** 
\noindent\textbf{Experimental Setup.}
%------------------------------------------
We evaluate \PUL{} both in a PIM and an in-storage NDP architecture. The PIM experiments are conducted on an Intel Xeon Silver 4110 CPU equipped with a PIM-enabled UPMEM memory PIM-DIMM (Codename P21). NDP experiments are conducted on an ARM Neoverse N1 System Development Platform \cite{N1SPD} equipped with 4x ARM N1-CPU cores. A Xilinx Alveo U280 FPGA (AU280) card serves as a smart/computational storage device and is connected via PCIe Gen3 x16 to the host. NDP processing is offloaded onto a PE-array, employing 14 MicroBlaze\cite{MicroBlaze:2022} soft cores (64bit) operating at 150 MHz. \PUL{} HW-modules are attached to each PE, providing up to 64KiB scratchpad memory (BRAM). In addition, a configurable HW-module (NVMulator\cite{Tamimi:NVMulator:ARC:2023}) emulates a variety of read and write latencies over conventional DRAM is part of the HW-design. For NVM emulation, we employ latencies 350ns read, 170ns write based on \cite{Leis:NVMlatencies:DAMON:2019, Gokhale:NVMlatencies:MEMSYS:2019, Swanson:NVMlatencies:arxiv:2019}. 

\noindent\textbf{Workoad and Dataset.}
%------------------------------------------
The \textit{workload} and the \textit{dataset} are based on the prefetching micro-benchmark from Kühn et al. \cite{Teubner:Prefetching:DAMON:2023}. The dataset is stored in an array and randomly accessed by a pre-generated trace. A transfer size of 64B is used if not stated otherwise. In NDP the preload-distance is configured to 64 if not explicitly stated in the experiments.

\noindent\textbf{Experiment 1: \PUL{} consistently improves compute efficiency.}
%------------------------------------------
\begin{figure}[t]
	\begin{center}
         \includegraphics[width=\columnwidth]{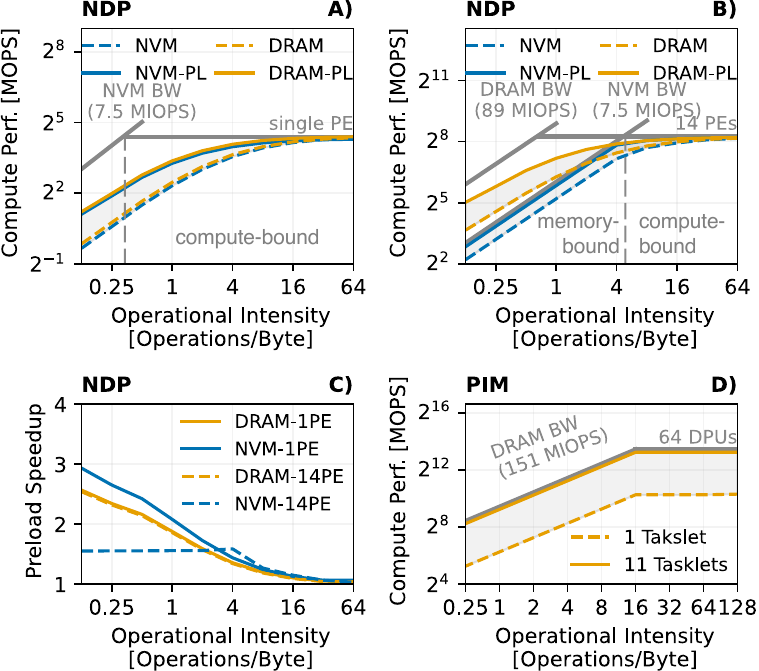}
         \caption{Roofline analysis of pre-loading and different operational intensities. \PUL{} achieves speedups through compute/IO interleaving in NDP and PIM on DRAM and NVM.}
	\label{fig:exp_1}
	\end{center} 
\end{figure} 
The performance of systems with weak PEs and low clock-frequencies highly depends on their ability to efficiently utilize available compute cycles. In fact, even with fast memories I/O wait times waste cycles and reduce overall system performance. However, slower memories such as NVM or Flash exacerbate matters, highlighting the need for software-based preloading.

To quantify the benefits of preloading (PL) in both PIM and NDP architectures, we investigate random cacheline-sized data transfers, while varying algorithmic/operational intensities (Fig. \ref{fig:exp_1}) to cover a wide range of DB operations. We compare the impact of DRAM and NVM latencies and evaluate the performance under two scenarios: (1) separate I/O and compute phases (no PL or 1 Tasklet), and (2) interleaved I/O and compute (with PL or 11 Tasklets).

Preloading consistently improves performance in NDP for operations with low algorithmic intensities, effectively hiding the NVM and DRAM access latency. As a result, \PUL{} achieves the \textit{same} compute performance with NVM and DRAM \textit{despite} the access latency gap of 3.5$\times$ (Fig. \ref{fig:headpicture}). Noticeably, higher NVM-latencies offer a greater potential for compute/IO interleaving with preloading, yielding higher speedups of up to 2.9$\times$ compared to DRAM's 2.5$\times$ (Fig. \ref{fig:exp_1}-C). The results also reveal that preloading with 14 PEs saturates the available NVM bandwidth, but still delivers up to 1.5$\times$ performance gain. In contrast, preloading on DRAM maintains its improvements with 14 PEs across all intensities as the high available bandwidth and throughput are being underutilized due to the slow individual PE compute performance (Fig. \ref{fig:exp_1}-B). In PIM settings we observe a different behavior, yielding speedups of 7.8$\times$ with low and 11$\times$ with high operational intensity (Fig. \ref{fig:exp_1}-D). Notably, the maximum PIM compute performance strongly correlates with the utilization of at least 11 Tasklets, which is not achievable in lower intensities due to bandwidth saturation. 

\noindent\textbf{Insights:}
The preload speedups achieved by \PUL{} align closely with those attained by software prefetching on common CPU architectures. However, newer architectures, like AMD Zen4 or Intel Sapphire Rapids, demonstrate that software prefetching is becoming more advantageous, with speedups of 4$\times$ to 5$\times$ \cite{Teubner:Prefetching:DAMON:2023}.
Yet, \PUL{} offers advantages in terms of simpler interfaces, avoidance of proprietary instructions, and obliviousness to cache-hierarchy specifics.

\noindent\textbf{Experiment 2: Interleaving DB operations and I/O allows for higher computational intensity at constant execution time.}
%------------------------------------------
Slower memories like NVM or Flash  significantly increase the I/O wait time, opening up the opportunity to interleave operation execution with I/O requests. As a result, higher computational intensity can be tolerated, while the execution time remains constant.

\begin{figure}[t]
	\begin{center}
         \includegraphics[width=\columnwidth]{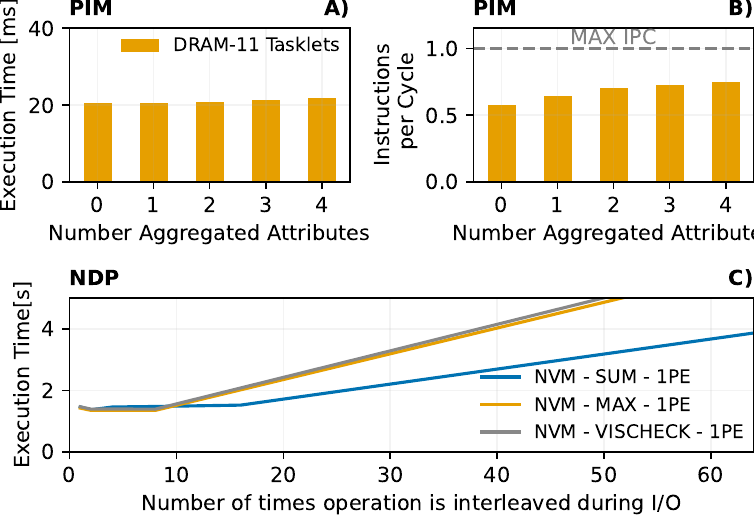}
		\caption{\PUL{} interleaves common DB-operations with data transfers without impacting execution times.}
	\label{fig:exp_2}
	\end{center} 
\end{figure}

To understand the intricate trade-offs between optimal compute and I/O-latency in preloading, we investigate aggregations in PIM. We increase the operational complexity by increasing the number of attributes being aggregated, thereby increasing the amount of computation, but keeping I/O-latency constant as attributes are stored row-wise and accessed in a single transfer. We measure both the execution time (Fig. \ref{fig:exp_2}-A) and the instructions per cycle (Fig. \ref{fig:exp_2}-B) to determine the optimal balance. Furthermore, we assess the compute intensity of common DB operations in NDP and position them in interleaving space on NVM (Fig. \ref{fig:exp_2}-C).

Our experimental results demonstrate that simply aggregating a single attribute in PIM falls short of efficiently interleaving compute and I/O, resulting in a low IPC value of 0.58 (max. IPC 1.0). Furthermore, we observe that increasing the number of attributes aggregated has minimal impact on execution time, but rather yields improvement in IPC values (Fig. \ref{fig:exp_2}-B). The analysis of wait times in NDP reveals that they are significantly higher than the compute costs associated with common DB operations, such as aggregations and MVCC visibility checking. These can be interleaved multiple times during a single data request (Fig. \ref{fig:exp_2}-C), highlighting the need to improve PE utilization by interleaving more compute.

\noindent\textbf{Insights:}
Attaining optimal I/O and PE utilization with \PUL{} mandates a delicate balance between compute-latency and I/O-latency. 
Higher memory latency, while potentially beneficial for interleaving compute-intensive operations, is challenging for data-intensive operations. Their low operational intensity and inherent data dependencies lead to poor PE utilization.
Increases in PE clock frequency aggravates matters. However, interleaving \textit{multiple} I/O requests increases I/O throughput utilization, improving PE utilization during data-intensive operations as we demonstrate next.

\noindent\textbf{Experiment 3: Preload distance.}
%------------------------------------------
\begin{figure}[t]
	\begin{center}
         \includegraphics[width=\columnwidth]{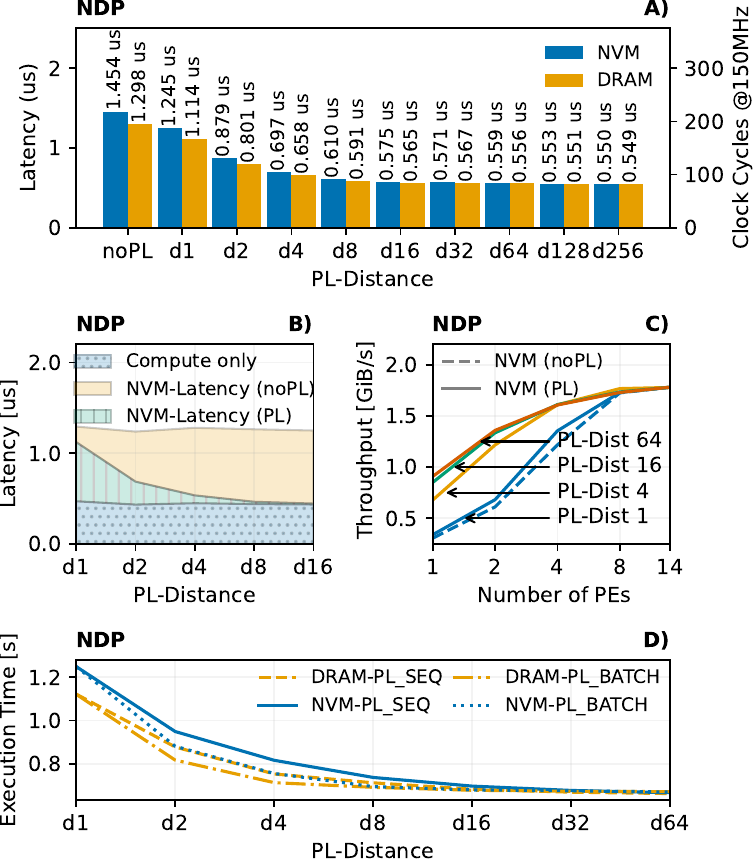}
		\caption{Interleaving multiple I/O-requests with \PUL{} can mitigate both NVM and DDR access latency, increase memory throughput, and maximize PE utilization.}
	\label{fig:exp_3}
	\end{center} 
  \vspace{-15pt}
\end{figure}  
The \textit{temporal preload window} between a data request and actual data consumption from scratchpad has a profound impact on preload efficiency. As the preload distance increases, the preload window for requesting data and ensuring timely availability expands, allowing higher flexibility in request scheduling, latency hiding, and an overall improvement in I/O throughput. Listing \ref{lst:preloading} shows an example with a preload distance of 4, preloading the fourth element in advance of computing on the zeroth element.

To evaluate the effect of preload distance, we incrementally increase the preload distance, while keeping the operational intensity constant and low by employing a \texttt{SUM} operation.
Our results (Fig. \ref{fig:exp_3}-A) demonstrate a continuous reduction in operation execution times as preload distance increases. Notably, it reaches a plateau at \texttt{d16}, Fig. \ref{fig:exp_3}-A, indicating that memory latency is fully mitigated and longer distances yield diminishing returns, which is confirmed by a separate examination of both the compute and I/O-latency (Fig. \ref{fig:exp_3}-B). Conversely, the reduction in overall latency through preloading improves I/O throughput and PE utilization (Fig. \ref{fig:exp_3}-C). 

Next, we investigate two distinct preload-strategies (Fig. \ref{fig:exp_3}-D). First, we explore sequential interleaving, 
where alternating preload requests and compute operations are overlapped (e.g., \texttt{PL[4]}$\rightarrow$ \texttt{compute[0]}$\rightarrow$\texttt{PL[5]}$\rightarrow$\texttt{compute[1]}\dots).  
Secondly, we examine batch-wise execution, wherein multiple requests (number corresponds to the distance) are sent first, followed by processing the data of the previous preload-batch. Our results indicate that batch-wise execution outperforms the sequential interleaving strategy in terms of overall I/O throughput, eliminating the computational delays intrinsic to the sequential approach. Notably, as the preload distance begins to fully compensate the memory latency (>\texttt{d16}), both methods provide similar performance, indicating that a batch-wise execution is beneficial for I/O interleaving while it has no effect on compute interleaving.

\noindent\textbf{Insights:}
By leveraging deep I/O request queues, our system can accommodate a high number of parallel pending data requests, outperforming the capabilities of traditional host-side prefetching (HW dependent) and PIM (limited by the number of Tasklets), overcoming the limitations reported in \cite{Teubner:Prefetching:DAMON:2023}. Moreover, preloading far into the future allows for optimal PE utilization, albeit constrained by available bandwidth and compute resources. 
Noticeably, in contrast to CPU-side cache-based prefetching methods, where excessive prefetch distances and cache contention can lead to premature eviction of prefetched data \cite{Teubner:Prefetching:DAMON:2023}, \PUL{}'s non-shared, explicitly managed scratchpads minimize this risk and do not require deliberate timing as discussed in \cite{aptget}. 
In conclusion, practical preload distances are short (<\texttt{d16}), require small scratchpad space and ensure high utilization of the weak PEs, while hiding latencies. Furthermore, by supporting longer preload distances \PUL{} become robust against varying operational  intensities.

\noindent\textbf{Experiment 4: Configurable transfer sizes.}
%------------------------------------------
\begin{figure}[t]
	\begin{center}
         \includegraphics[width=\columnwidth]{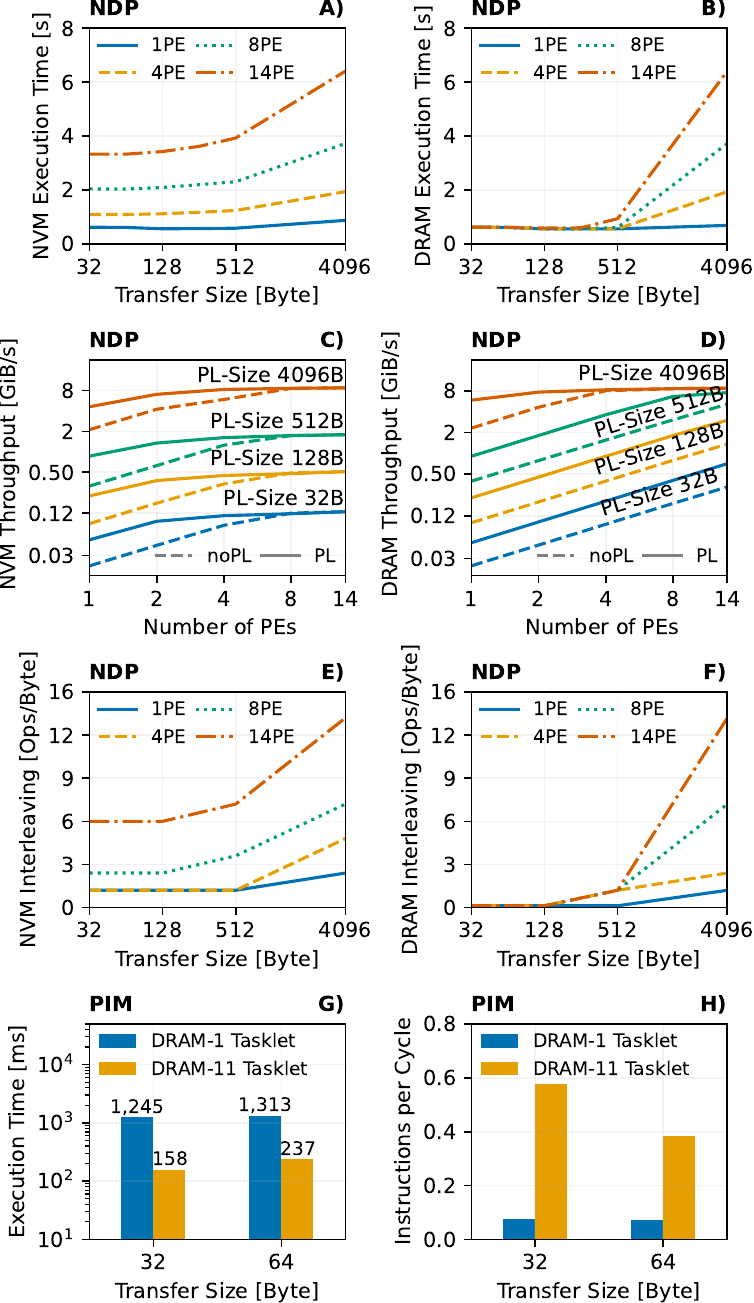}
		\caption{ \PUL{} greatly reduces the number of PEs required to fully utilize bandwidth while efficiently interleaving larger transfers to increase throughput. }
	\label{fig:exp_4}
	\end{center} 
  \vspace{-10pt}
\end{figure}
In this experiment we investigate two claims. First, while CPU-based prefetchers are limited by a fixed cacheline-sized granularity, which drives many layout and access optimizations, \PUL{} overcomes the limitation with configurable transfer sizes, significantly improving performance. Secondly, leveraging larger transfers, \PUL{} achieves high memory bandwidth, exceeding that of CPU-based software prefetchers.

While a single large data transfer results in a higher latency compared to a smaller one, multiple sequential transfers have worse total latency compared to the larger one, due to lower request management overhead on the PE. This effect is particularly pronounced in memory architectures lacking HW prefetchers and complex caching hierarchies. The higher latencies of larger transfers can, however, be effectively hidden through \PUL{} I/O-interleaving. \PUL{} increases bandwidth and reduces the amount of PE resources required to fully utilize the memory bandwidth. In addition, the higher access latency associated with larger transfers offer the potential for more compute interleaving.

To investigate the effects of transfer sizes on execution time, memory bandwidth, and I/O-interleaving, we conduct multiple experiments, in which we execute the same sequence of requests per PE with varying transfer sizes, while maintaining minimal compute-overhead through the use of a simple \texttt{SUM} operation. Furthermore, we increase the algorithmic intensity by varying the number of values for the \texttt{SUM} until further interleaving of computations becomes impossible and report this compute interleaving potential for increasing transfer sizes. In addition, varying the number of PEs allocated during execution allows us to determine how much resources are required to fully utilize the bandwidth with and without \PUL{}. Increasing the number of PEs even further naturally increases bandwidth-boundness, allowing us to evaluate its impact on \PUL{}.

In the case of NVM (Fig. \ref{fig:exp_4}-A), we observe that a single PE can interleave larger transfer sizes, up to 512B, without affecting execution time. Furthermore, we find that increasing the number of allocated PEs leads to an increase in execution time, as the system quickly reaches bandwidth saturation. This also effects I/O interleaving efficiency, as larger transfers increase pressure on memory bandwidth, resulting in worse execution times. In terms of memory throughput (Fig. \ref{fig:exp_4}-C), we observe that \PUL{} is able to saturate bandwidth by leveraging 2-3 PEs, whereas without \PUL{}, an equivalent level of throughput can only be reached using at least 8 PEs. As expected, the transfer size greatly improves memory bandwidth utilization. However, the system throughput is saturated at transfer sizes of 4096B, preventing us from reaching speeds exceeding 8 GiB/s. In comparison, DRAM's much higher throughput allows \PUL{} to perfectly interleave requests up to 512B, while utilizing the maximum number of PEs (Fig. \ref{fig:exp_4}-B). Yet, DRAM is also constrained by our current system, also limiting the bandwidth to 8 GiB/s (Fig. \ref{fig:exp_4}-D). While we do not observe significant impact on compute interleaving at larger transfer sizes without saturating the NVM bandwidth (Fig. \ref{fig:exp_4}-E), with DRAM we do (Fig. \ref{fig:exp_4}-F), beyond 128B. These experiments also reveal a notable correlation between the degree of bandwidth-boundness and its direct impact on the overall prefetch latency, as well as a related effect on reduced PE-utilization with higher potential for compute interleaving. 

We also evaluate the impact of transfer sizes on PIM execution times (Fig. \ref{fig:exp_4}-G) and PE utilization (Fig. \ref{fig:exp_4}-H). PIM PEs immediately become bandwidth-bound executing a simple \texttt{SUM}, resulting in an IPC value of 0.58 utilizing 32B transfers. Increasing the transfer size further has a detrimental effect on performance, leading to higher access latencies, worse execution times, and lower IPC values as the additional data is not sufficiently amortized by compute.

\noindent\textbf{Insights:}
Compared to fixed cacheline-sized CPU-prefetchers, configurable transfer sizes in \PUL{} allow data requests to be more easily optimized and aligned to a variety of data layouts/structures and algorithms, while efficiently supporting compute and I/O interleaving with less PE resources. These benefits cannot be achieved on traditional CPUs due to the limitations imposed by unpredictable cache evictions and cache contention.

\noindent\textbf{Experiment 5: Unloading.}
%------------------------------------------
Materializing final or temporary PE results incurs a problem inverse to preloading, i.e., PEs are slowed down by expensive flushes from scratchpad memory to persistent storage. Unloading enables efficient interleaving of those flushes, which can occur either in several small random locations, or sequential blocks. 

\begin{figure}[t]
	\begin{center}
         \includegraphics[width=\columnwidth]{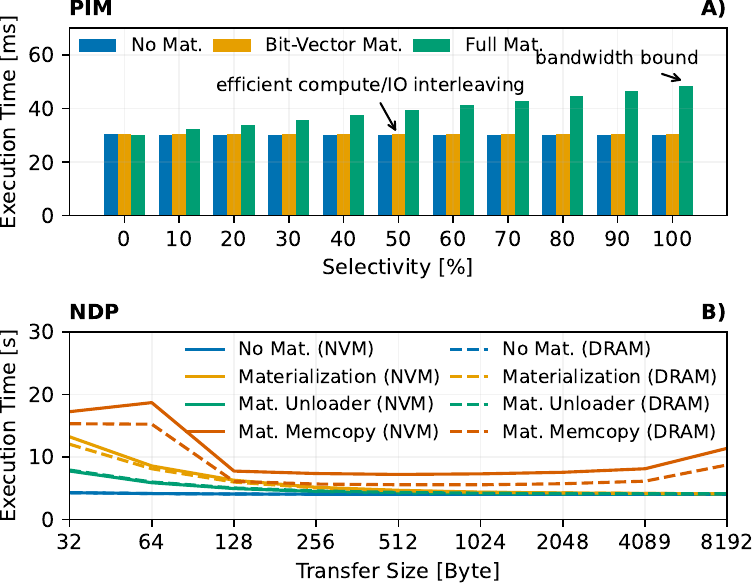}
		\caption{Leveraging \PUL{}'s unloading capabilities during result materialization provide equally important improvements to preloading.}
	\label{fig:exp_5}
	\end{center} 
 \vspace{-10pt}
\end{figure}

To demonstrate the effectiveness of I/O interleaving in \textit{unloading} settings, we offload a simple filter operation onto PIM and measure different selectivities. When records are not filtered out, we materialize them in sequential blocks, representing the result-set. Our results show that fully materializing the whole record is not compute intensive, but has a significant impact on the bandwidth and, due to the already bandwidth-bound filter operation, leads to an increasingly worse performance (Fig. \ref{fig:exp_5}-A). By introducing a positional bit-vector to encode addresses of filtered records instead, we reduce the required bandwidth while also optimizing PE utilization by interleaving the addition compute for the bit-vector creation. This allows us to fully mitigate the materialization overhead.

We also investigate the impact of unloading in NDP and analyze the effects of varying transfer sizes. We iteratively update values stored in the scratchpad and periodically trigger flushing, when a threshold transfer size is exceeded. Our results (Fig. \ref{fig:exp_5}-B) compare the execution time of materialization leveraging \PUL{} with a baseline excluding \PUL{}. We also include numbers for standard \texttt{memcpy}, which however falls back to cache-line sized transfers.

Flushing larger data sizes benefits most materialization strategies due to the reduction on flush management overheads and better PE utilization, with the exception of memcpy, which at 2048b incurs too many requests, saturating bandwidth. \PUL{} provides the best improvement with frequent random flushes that are also accompanied by low operational intensity. This is consistent with our previous experimental findings.

\noindent\textbf{Insights:}
The exclusive PE managed scratchpad enables the unloading of large blocks of data at once, overcoming the limitations of cache-line write back (CLWB) instructions on the CPU. 
Compared to preloading, where timely availability is crucial due to data dependencies, persisting results often have relaxed timing constraints, as they only require periodic flushing until the end of execution. However, when synchronization mechanisms like locking and updating crucial data structures like indices are involved, ensuring that all necessary data is persisted before releasing locks becomes critical. Leaveraging \PUL{}'s explicit software synchronization mechanisms makes sure that data is persisted in such cases.

\section{Conclusions}
\label{sect:conclusions}
%*****************************************
In this paper, we presented \PUL{} and a custom DMA-engine to address the challenges posed by post-Moore architectures such as PIM and NDP. Notably, the cache/memory organization in these systems differ from traditional CPUs as they rely on scratchpad memory and explicitly controlled DMA-engines, making widely adopted cache optimizations techniques ineffective, while still operating on the same data and layouts. \PUL{}'s flexibility in interleaving compute and I/O via software significantly improves compute performance and resource utilization, aligned with software prefetching techniques on common CPUs, while providing developer-friendly interfaces decoupled from vendor specific knowledge.

\begin{acks}
%*****************************************
This work has been partially supported by  \emph{DFG Grant neoDBMS.2 -- 419942270}.
\end{acks}
%*****************************************

\balance

\newpage
% \bibliographystyle{abbrv}
% \bibliography{references}
\bibliographystyle{ACM-Reference-Format}
\bibliography{references}

% ****************** APPENDIX **************************************
%\begin{appendix}
%\dots 
%
%
%\end{appendix}

\end{document}